\documentclass[a4paper,11pt]{article}
\pdfoutput=1
\usepackage{jcappub}

\usepackage{amsmath}
\usepackage{bm}
\usepackage{xcolor,color,colortbl}
\usepackage[utf8]{inputenc}
\usepackage{hyperref}
\usepackage{url}
\usepackage{graphicx}
\usepackage{multirow,bigdelim}
\usepackage{amssymb}
\usepackage[amssymb]{SIunits}
\usepackage{multirow}
\usepackage{bbold}
\usepackage{verbatim}

\definecolor{darkgreen}{HTML}{339933}

\usepackage{scalerel,tikz}
\usetikzlibrary{svg.path}
\definecolor{orcidlogocol}{HTML}{A6CE39}
\tikzset{orcidlogo/.pic={
 \fill[orcidlogocol] svg{M256,128c0,70.7-57.3,128-128,128C57.3,256,0,198.7,0,128C0,57.3,57.3,0,128,0C198.7,0,256,57.3,256,128z};
 \fill[white] svg{M86.3,186.2H70.9V79.1h15.4v48.4V186.2z}
 svg{M108.9,79.1h41.6c39.6,0,57,28.3,57,53.6c0,27.5-21.5,53.6-56.8,53.6h-41.8V79.1z M124.3,172.4h24.5c34.9,0,42.9-26.5,42.9-39.7c0-21.5-13.7-39.7-43.7-39.7h-23.7V172.4z}
 svg{M88.7,56.8c0,5.5-4.5,10.1-10.1,10.1c-5.6,0-10.1-4.6-10.1-10.1c0-5.6,4.5-10.1,10.1-10.1C84.2,46.7,88.7,51.3,88.7,56.8z};
}}
\newcommand\orcidicon[1]{\href{https://orcid.org/#1}{\mbox{\scalerel*{
\begin{tikzpicture}[yscale=-1,transform shape]
\pic{orcidlogo};
\end{tikzpicture}
}{|}}}}

\title{Harmonic analysis of discrete tracers of large-scale structure}

\author[]{Antón Baleato Lizancos~\orcidicon{0000-0002-0232-6480}}
\author[]{and Martin White~\orcidicon{0000-0001-9912-5070}}
\affiliation{Berkeley Center for Cosmological Physics, UC Berkeley, CA 94720, USA}
\affiliation{Department of Physics, University of California, Berkeley, CA 94720, USA}
\affiliation{Lawrence Berkeley National Laboratory, One Cyclotron Road, Berkeley, CA 94720, USA}

\emailAdd{a.baleatolizancos@berkeley.edu}
\emailAdd{mwhite@berkeley.edu}

\abstract{It is commonplace in cosmology to analyze fields projected onto the celestial sphere, and in particular density fields that are defined by a set of points e.g.\ galaxies.  When performing an harmonic-space analysis of such data (e.g.\ an angular power spectrum) using a pixelized map one has to deal with aliasing of small-scale power and pixel window functions.   We compare and contrast the approaches to this problem taken in the cosmic microwave background and large-scale structure communities, and advocate for a direct approach that avoids pixelization.  We describe a method for performing a pseudo-spectrum analysis of a galaxy data set and show that it can be implemented efficiently using well-known algorithms for special functions that are suited to acceleration by graphics processing units (GPUs).  The method returns the same spectra as the more traditional map-based approach if in the latter the number of pixels is taken to be sufficiently large and the mask is well sampled.  The method is readily generalizable to cross-spectra and higher-order functions.  It also provides a convenient route for distributing the information in a galaxy catalog directly in harmonic space, as a complement to releasing the configuration-space positions and weights, and a route to spectral apodization. We make public a code enabling the application of our method to existing and upcoming datasets. } 

\begin{document}
\maketitle
\flushbottom

\section{Introduction}
\label{sec:introduction}

Observational cosmology is entering an exciting era of ``big surveys'', each of which will map large fractions of the sky with unprecedented depth and fidelity.  Examples include DESI \cite{DESI}, Euclid \cite{laureijs2011euclid,EUCLID18}, SPHEREx \cite{Dore14}, LSST on the Vera Rubin Observatory \cite{LSST}, Roman \cite{Roman}, the Simons Observatory \cite{SimonsObs}, the South Pole Observatory and CMB-S4 \cite{CMBS4}.  These surveys probe the Universe in complementary ways, and the correlated signals within the datasets they will produce will offer a wealth of insight into questions of cosmology, astrophysics and fundamental physics \cite{SnowmassCF}.  In this paper we are concerned with techniques for performing auto- and cross-correlations of scalar fields on the sphere, which includes photometric galaxy clustering, CMB lensing, the thermal and kinematic Sunyaev-Zeldovich effects \cite{Sunyaev80,Carlstrom02} and many other probes.  Our particular focus will be on how to handle galaxy data, which is provided as a set of points with significant amounts of small-scale power (e.g.\ shot noise), in a harmonic-space analysis.

The CMB community has developed sets of tools for dealing with pixelized maps of continuous fields defined on the sphere (e.g.~CMB temperature or polarization and lensing convergence), with the most common being the Healpix pixelization scheme \cite{Gorski05} and the pseudo-$C_\ell$ method for measuring their (angular) power spectra \cite{Hivon02}.  These methods work very well for spectra without large amounts of small-scale power\footnote{CMB experiments observing at microwave frequencies have beams much larger than typical optical surveys and thus smear out much more of the small-scale structures present in the sky.  In addition many of the signal components (e.g.\ primary CMB anisotropies) have intrinsically very red spectra.}, and very efficient ``community codes'' are available for many of the standard operations (e.g.~\texttt{NaMaster} \cite{Alonso19}, \texttt{PolSpice} \cite{chon_challinor_04}).  When dealing with projected galaxy data most authors have tried to take advantage of these tools, and so as a first step have produced pixelized maps of the (projected) galaxy overdensity (see e.g.\ ref.~\cite{Garcia21} for a pedagogical example from the recent literature). Typically (e.g.~\cite{Nicola20,Garcia21,White21}) these maps are created by binning the galaxy positions, with their systematics weights, into a map, and then doing the same for a random catalog that defines the sample in the absence of clustering\footnote{Instead of randoms, some authors (e.g., refs.~\cite{Garcia21,Nicola20}) work with `completeness weights' that define the local mean density as a modulation of the average number of galaxies per pixel by the fraction of each pixel that is rendered unobservable due to non-idealities. The randoms serve the same purpose while providing a convenient avenue to convert between the various pixelizations and coordinate systems involved in the calculation of completeness. Due to this equivalence, we will not address completeness weights explicitly in what follows.}.  The ratio of the two maps then defines the overdensity field, $1+\delta_{\rm gal}$.  However, this procedure has some drawbacks.   First of all, the tails of the distribution of any inaccuracies in the random (or mask) map are enhanced by this division process, which introduces a non-linear function of e.g.~the systematics weights. Another issue is that because of shot-noise the galaxy field has significant small-scale power which means that aliasing is a concern.  Aliasing could be reduced by interlacing or a higher-order charge assignment scheme \cite{Hockney88,Jing04,Sefusatti15}, but this is difficult in curvilinear coordinates and common practice is to assign galaxies and randoms to a single pixel.  The finite size of the pixels also means that the pixel window function can depart from unity at the angular scales targeted by analyses; moreover, when these scales are close to the pixel size, the non-trivial shapes of the pixels become important and pixelization corrections become computationally intractable\footnote{Pixelized observations result from averaging the underlying signal within each pixel -- i.e., convolving the signal with a `pixel window function'. On scales much larger than the pixel size, pixels can be assumed to all be the same shape, and the impact of pixelization is well captured by a simple filtering of the angular spectra by a function that depends only on the angular multipole $\ell$; see, e.g., ref.~\cite{hivon_e_pixel_nodate}. An analysis that fully captures the non-trivial shapes of the pixels is computationally intractable.}.  For all these reasons the practice has been to employ a high resolution map (e.g.\ ref.~\cite{Krolewski20} and the discussion in the appendices of refs.~\cite{Leistedt13,Alonso19}). However, fine pixelizations come with a computational cost. Since the density field involves dividing by the random map, a large number of random points is needed in order to prevent stochastic noise in the randoms from significantly distorting the statistics of the overdensity field. And, perhaps more importantly, the spherical harmonic transform algorithm scales with the number of pixels -- even in the highly efficient \texttt{HealPix} scheme, it goes as $N_{\rm pix}^{3/2}$~\cite{Gorski05}. 

In this paper we advocate for a different approach motivated by the procedure that is commonly employed in analyzing three dimensional galaxy (redshift) surveys (\cite{Feldman94}; hereafter FKP).  The ``FKP approach'' defines the power spectrum to be the square of the Fourier transform of a normalized difference between the data and a scaled set of randoms.  Thus, if $\alpha$ is the ratio of the number of data to the number of randoms, then the FKP field is proportional to $n_g(\bm{x})-\alpha n_r(\bm{x})$, with the proportionality constant chosen for convenience.  Since the randoms define a mean density, $\bar{n}$, and the data define $\bar{n}(1+\delta)$, the difference is $\bar{n}\,\delta$.  The square of the Fourier transform of this field has as expectation value the power spectrum convolved with a window function.  If we were to follow this approach, the standard definition of the overdensity, $\delta (\bm{x}) \propto n_g(\bm{x})/n_r(\bm{x})-1$, a non-linear operation involving division by the randoms or mean density, would be replaced with $\delta (\bm{x}) \propto n_g(\bm{x})-\alpha n_r(\bm{x})$. Since the latter is linear in data and randoms, the harmonic-space overdensity could then be obtained directly by subtracting the spherical harmonic transforms of data and scaled randoms. In the limit that the objects are pointlike, the spherical harmonic transform of these fields is simply a sum of spherical harmonics evaluated as the positions of the objects, $Y_{\ell m}^{\star}(\theta_i,\phi_i)$.  This can be done without recourse to a map or pixelization, thus avoiding the pixel window function and any aliasing. Highly efficient routines for computing $Y_{\ell m}$ exist that make this appealing and allow us to focus computational effort on the particular $\ell$s which are of interest. Moreover, the key steps of the computation are well-suited for acceleration on Graphics Processing Units (GPUs). Finally, this approach provides a convenient method for distributing the galaxy information in harmonic space.  The coefficients, $a_{\ell m}$, of the harmonic expansion can be released directly (alongside catalogs of positions and weights in configuration space).  Rotating coordinates or generating a pixelized map from these coefficients is straightforward using existing tools\footnote{e.g.\ https://healpy.readthedocs.io/en/latest/} \cite{Gorski05}.

The outline of the paper is as follows.  In \S\ref{sec:algorithm} we define the basic algorithm, focusing on the case of 2-point statistics (i.e.\ the angular auto-power spectrum) for definiteness though the generalization to cross-spectra and to higher order functions is reasonably straightforward. Then, in \S\ref{sec:examples}, we consider two examples that illustrate the benefits of our proposed approach. We conclude with a general discussion in \S\ref{sec:conclusions}. Some implementation details are relegated to the Appendices.

\section{Power spectrum algorithm}
\label{sec:algorithm}

Let us now explain in detail the proposed procedure, which closely follows the ``pseudo-$C_\ell$'' method \cite{Hivon02} and the FKP method\footnote{Ref.~\cite{Feldman94} additionally describe ``optimal weights'' -- frequently referred to as ``FKP weights'' -- that balance sample variance and shot noise.  We shall not consider those here.} \cite{Feldman94}. A high-level summary is provided in \S\ref{sec:algorithm_summary}.

\subsection{Pseudo-spectra}

Given either a set of weighted randoms or a pixelized map describing the survey ``mask'' or ``window'', which we denote $w(\theta,\phi)$, let $w_{\ell m}$ be this window's spherical harmonic transform (SHT; we describe how to compute $w_{\ell m}$ later).  In large-scale structure studies this would typically be denoted $\bar{n}$, the mean density in the absence of clustering, but we shall follow the CMB convention where this is a ``window''.  Define the square of $w_{\ell m}$ to be a window function\footnote{If $W_\ell$ is estimated by a small number of randoms, one should subtract the shot-noise contribution.}
\begin{equation}
    W_\ell = \frac{1}{2\ell+1}\sum_{m=-\ell}^\ell \left| w_{\ell m} \right|^2\,.
\label{eqn:wldef}
\end{equation}
The masked/window-convolved data minus scaled randoms or normalized mask is $w(1+\delta)-w=w\delta$ and its SHT is the difference of the SHTs of each term.  This product has SHT~\cite{baleato_lizancos_impact_2023}
\begin{align}
    \left\{w\, \delta\right\}_{\ell m} \equiv   (-1)^{m} \sum_{\ell _1 m_1} \sum_{\ell _2 m_2} G^{\ell \ell_1\ell_2}_{-m m_1 m_2} w_{\ell _1 m_1} \delta_{\ell _2 m_2}\,,
\end{align}
where we have used the definition of the Gaunt integral
\begin{align}
    G^{\ell_1\ell_2\ell_3}_{m_1 m_2 m_3} &\equiv \int d\hat{n}\ Y_{\ell _1 m_1}(\hat{n}) Y_{\ell _2 m_2}(\hat{n}) Y_{\ell _3 m_3}(\hat{n}) \nonumber \\
    & = \sqrt{\frac{(2\ell_1+1)(2\ell_2+1)(2\ell_3+1)}{4\pi}}\begin{pmatrix}\ell_1&\ell_2&\ell_3 \\ m_1&m_2&m_3 \end{pmatrix} \begin{pmatrix}\ell_1&\ell_2&\ell_3 \\ 0&0&0 \end{pmatrix} \,,
\label{eqn:gaunt}
\end{align}
with the terms in parentheses indicating $3j$ symbols \cite{Edmonds96} (see Appendix \ref{app:3j} for an algorithm to compute them efficiently).
Now average over $m$ to define
\begin{equation}
    \widehat{C}_\ell \equiv \frac{1}{2\ell+1}\sum_{m=-\ell}^{\ell}
    \left| \vphantom{I} \left\{w\, \delta\right\}_{\ell m} \right|^2
    \quad .
\label{eqn:pseudoCl}
\end{equation}
Taking the expectation value
\begin{align}
    \left\langle \left| \vphantom{I} \left\{w\, \delta\right\}_{\ell m} \right|^2 \right\rangle
    &= \sum_{\ell_1 \cdots \ell_4}\sum_{m_1 \cdots m_4} G_{-mm_1m_2}^{\ell\ell_1\ell_2} G_{-mm_3m_4}^{\ell\ell_3\ell_4}
       w_{\ell_1m_1}w_{\ell_3m_3}^\star \left\langle \delta_{\ell_2m_2}\delta_{\ell_4m_4}^\star\right\rangle \\
    &= \sum_{\ell_1 \cdots \ell_3}\sum_{m_1 \cdots m_3} G_{-mm_2m_1}^{\ell\ell_2\ell_1} G_{-mm_2m_3}^{\ell\ell_2\ell_3}
       w_{\ell_1m_1}w_{\ell_3m_3}^\star C_{\ell_2}
\end{align}
and using \cite{Edmonds96}
\begin{align}
    \sum_{m_4 m_1} G^{\ell_4 \ell_1 \ell_2}_{-m_4 m_1 m_2} G^{\ell_4 \ell_1 \ell_3}_{-m_4 m_2 m_3} 
    &= \delta^{(K)}_{\ell_2 \ell_3} \delta^{(K)}_{m_2 m_3}\frac{(2\ell_4+1)(2\ell_1+1)}{4\pi}
    \begin{pmatrix} \ell_4 & \ell_1 & \ell_2 \\ 0&0&0 \end{pmatrix}^2\,,
\end{align}
where $\delta^{(K)}$ is the Kronecker delta function, we find \cite{Hivon02}
\begin{equation}
    \left\langle \widehat{C}_\ell \right\rangle
    = \sum_{\ell'} M_{\ell\ell'} C_{\ell'}
    \quad \mathrm{with} \quad
    M_{\ell\ell'} = \frac{2\ell'+1}{4\pi} \sum_{\lambda} (2\lambda+1)
    \begin{pmatrix} \ell & \ell' & \lambda \\ 0&0&0 \end{pmatrix}^2
    W_{\lambda}\,,
\label{eqn:chat}
\end{equation}
where $C_{\ell'}$ is the `true' angular power spectrum of the overdensity field.

This mirrors the ``normal'' derivation of the pseudo-$C_\ell$ calculation and definition of the mode-coupling matrix, $M_{\ell\ell'}$ \cite{Hivon02}.  In fact, we can see that simply differencing the data and random catalogs achieves the same result as the more usual approach \cite{Nicola20} of generating an overdensity field $\delta(\hat{n})$ (by dividing a data catalog by a random catalog and subtracting one) multiplying it by the mask $w(\hat{n})$ (generated from the random catalog and possibly apodized), and transforming the product: $(D/R-1)\times R=D-R$.

This $\widehat{C}_\ell$ (likely binned into bandpowers) and $M_{\ell\ell'}$ are sufficient to enable a comparison of theory and data. However, in general, the spectra are not reported as $\widehat{C}_\ell$.  Within the CMB community the normal procedure \cite{Hivon02,Nicola20} is to bin the $\hat{C}_\ell$ in bands of $\ell$ and apply a mode-decoupling.  The mode-decoupling step ensures that if the true theory were well approximated by a series of constant bandpowers in the chosen bins then the reported bandpowers would match this, i.e.\ the mode-decoupling step is a convenience.  On the other hand, the FKP procedure, commonly used in large-scale structure, does not apply such a step but simply a normalization\footnote{Ignoring the ``FKP weights'', the normalization adopted in ref.~\cite{Feldman94} divides by the square root of $\int dV\, \bar{n}^2$.}.

Let us consider each approach in turn.  In the standard pseudo-$C_\ell$ approach we take a set of bins,  $B$, defined through a matrix $\theta_b^\ell$ with $b\in B$.  A common choice would be $\theta_b^\ell=1/\Delta\ell$ for $\ell$ in bin $b$ and zero otherwise, where $\Delta\ell$ is the number of $\ell$s in bin $b$.  This simply averages the $\widehat{C}_\ell$ within each bin.  Alternatives to weight the $\ell$s within a bin differently can also be considered.  Now bin $\widehat{C}_\ell$ into these bins to form $\widehat{C}_b=\sum_\ell \theta_b^\ell\widehat{C}_\ell$.  Further assume the true $C_\ell$ is piecewise constant within the same bins, with values $C_b$. The binned coupling matrix, $M_{bb'}=\sum_{\ell\in b}\sum_{\ell'\in b'} \theta_b^\ell M_{\ell\ell'}$, relates $\langle\widehat{C}_b\rangle$ to $C_{b'}$ and is invertible for sufficiently broad bins.  Multiply by the inverse of this matrix so that
\begin{equation}
    \widetilde{C}_b = \sum_{b'} M_{bb'}^{-1} \widehat{C}_{b'}
    = \sum_{b'} M_{bb'}^{-1} \sum_{\ell m}
    \frac{\theta_{b'}^\ell}{2\ell+1}
    \left| \vphantom{\int} a_{\ell m}^{(d)}-w_{\ell m} \right|^2\,.
\label{eqn:mode_coupling}
\end{equation}
Note that we have defined our overdensity field following FKP, denoting the SHT of data and scaled randoms as $ a_{\ell m}^{(d)}$ and $w_{\ell m}$, respectively. The above `mode-decoupled' bandpowers have expectation value 
\begin{equation}
    \langle\widetilde{C}_b\rangle = \sum_{b'} M_{bb'}^{-1} \, \langle\widehat{C}_{b'}\rangle
    = \sum_{b'\ell\ell'} \left( M_{bb'}^{-1} \theta_{b'}^\ell M_{\ell\ell'}\right) \, C_{\ell'}
    \equiv \sum_{\ell'} \mathcal{M}_{b\ell'}\, C_{\ell'}\,.
\label{eqn:expectation_value}
\end{equation}
Implicitly, the mode-coupling matrix knows about the normalization of the weighted data (which matches that of the weighted randoms by construction), so the mode-decoupled bandpowers are independent of the chosen normalization.

The mode-coupling matrix $M_{\ell\ell'}$ is strictly positive.  However, since the mode-decoupling step involves $M_{bb'}^{-1}$, the matrix $\mathcal{M}_{b\ell'}$ for any bin $b$ depends upon all of the chosen bins and will typically have negative entries.  The reason is that it is trying to correct for changes in the shape of the pseudo-spectrum induced by the harmonic-space convolution with the mask\footnote{This can lead to odd behavior in bandpowers near the edges of the domain, as the mode-deconvolution algorithm implicitly assumes that the true $C_\ell$ are zero beyond the multipole range we have chosen to decouple (see also ref.~\cite{Singh21} for a related discussion).}.

An alternative approach, closer to how FKP proposed to process 3D power spectra, would be to simply normalize and bin $M_{\ell\ell'}$ to define a new pseudo-spectrum which would have only positive weights that do not depend on distant bins.  One possibility is
\begin{equation}
    \widetilde{C}'_b = \mathcal{N} \sum_{\ell} \theta_b^\ell \widehat{C}_{\ell}
    \quad\mathrm{with}\quad
    \left\langle \widetilde{C}'_b\right\rangle
    =  \mathcal{N}  \sum_{\ell\ell'} \theta_b^\ell M_{\ell\ell'} C_{\ell'}
     = \sum_{\ell'}\mathcal{M}'_{b\ell'} C_{\ell'}
    \quad .
\end{equation}
While the choice of $\mathcal{N}$ is arbitrary, a sensible choice is to leave the amplitude of a shot-noise spectrum unchanged.  Since $C_\ell$ is constant  for shot-noise, using the addition theorem for spherical harmonics and the conventional normalization $P_\ell(1)=1$ we have
\begin{equation}
  \langle \widehat{C}_\ell  \rangle 
  = \left[ \frac{\int d\Omega\ w^2(\hat{n}) }{4\pi} \right]\  C_\ell^{\rm shot}
  = \left[ \frac{W_0 \int d\Omega\ w^2(\hat{n}) }{\left| \int d\Omega\ w(\hat{n}) \right|^2} \right]
  \  C_\ell^{\rm shot}
  = \left[\sum_{\ell'} \frac{(2\ell'+1)}{4\pi} W_{\ell'} \right]\ C_\ell^{\rm shot}\,,
\end{equation}
where the last equality comes from expanding $w(\hat{n})$ in $Y_{\ell m}$ and using the orthonormality of the spherical harmonics. For the $\theta_b^\ell$ described above, choosing $\mathcal{N}$ as the inverse of the term(s) in square brackets leaves the amplitude of shot-noise at high $\ell$ unchanged. 

Note that this normalization does not address the change in the shape of the power spectrum induced by the convolution in Eq.~(\ref{eqn:chat}). The shape of the reported power spectrum will therefore differ from the true one ``on the sky'', especially at low $\ell$ where $W_\ell$ has significant support. However,  we emphasize that this is simply a convention for quoting the data: from the perspective of comparing theory and data one multiplies the theory curve by the supplied $N_{\rm bin}\times N_{\ell}$ matrix ($\mathcal{M}_{b\ell}$ or $\mathcal{M}'_{b\ell}$ in our notation above) to compare to the supplied data ($\widetilde{C}_b$ or $\widetilde{C}'_b$).  Any self-consistent convention is equivalent.

The resulting pseudo-spectrum includes stochastic or Poisson noise.  This could be subtracted if desired (as e.g.\ done in ref.~\cite{Feldman94}), though the approach gaining traction in the field is to include it in both the measurement and the model.  This then allows freedom for the noise to have a non-Poisson value whose degree of departure from the Poisson value can be handled with priors if so desired.

Finally we mention that if the computation of the pseudo-spectrum is being done as part of a Monte Carlo in which the footprint (and/or random catalog) is not changing, then the computation of the $w_{\ell m}$, $W_\ell$ and $M_{\ell\ell'}$ does not need to be redone.  In addition, the $Y_{\ell m}(\hat{n}_k)$ grid doesn't need to be recomputed if $\ell_{\rm max}$ is unchanged.  This can speed up the computation.

\begin{figure}
    \centering
    \resizebox{\columnwidth}{!}{\includegraphics{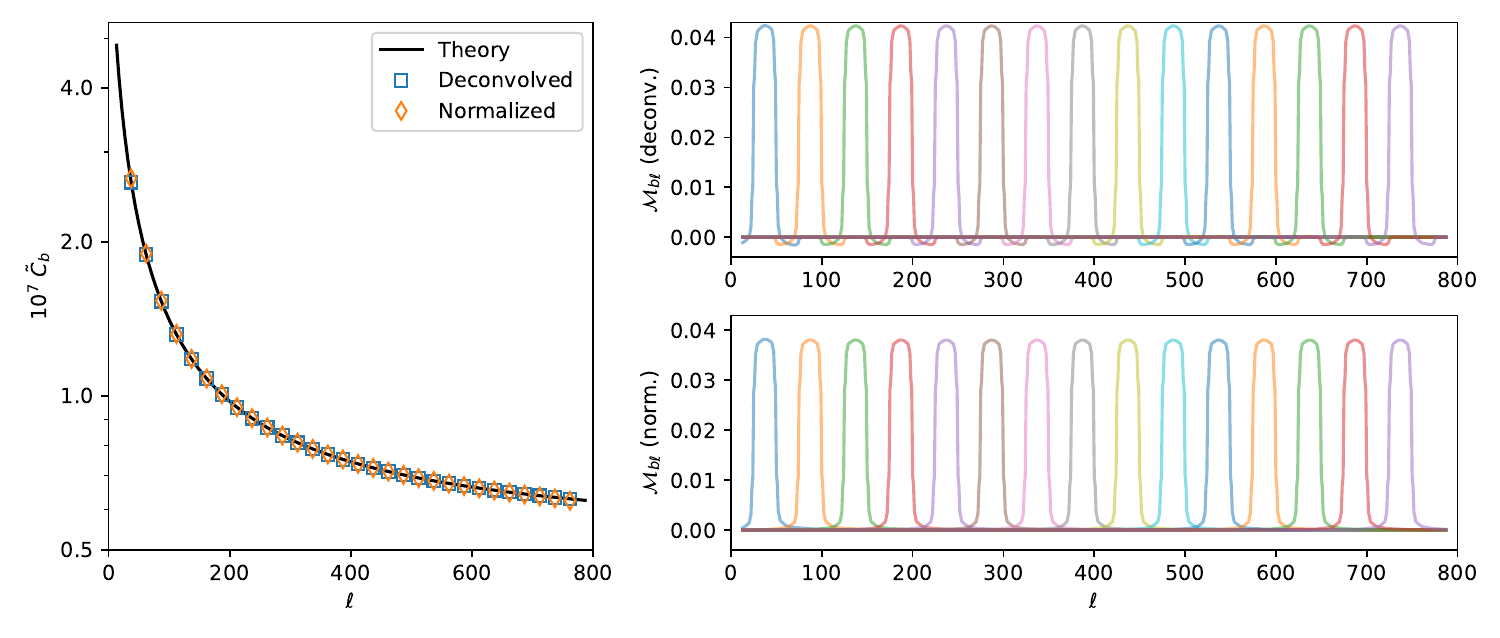}}
    \caption{A comparison of the two conventions for $\tilde{C}_b$ for the footprint shown in Fig.~\ref{fig:map} and linear bins of $\Delta\ell=25$.  (Left) The `true' full-sky spectrum is shown as the solid black line, the mode-decoupled spectrum as blue points and the normalized spectrum as orange points.  (Right) The matrices $\mathcal{M}_{b\ell}$ for each convention, shown for every other bandpower.
    }
    \label{fig:conventions}
\end{figure}

\subsection{Direct SHT}

\begin{figure}
    \centering
    \resizebox{0.9\columnwidth}{!}{\includegraphics{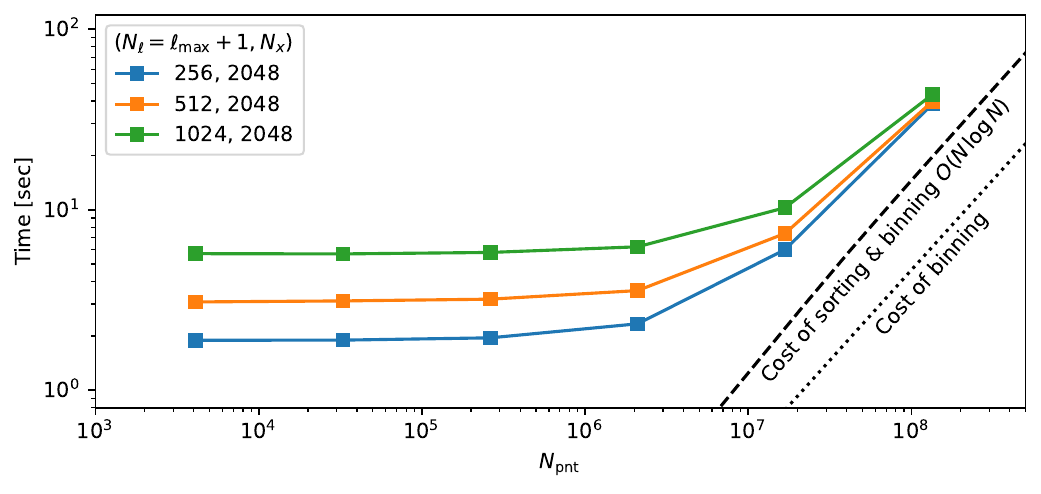}}
    \caption{Speed of the direct SHT, running on four NVIDIA A100 GPU cores of the Perlmutter computer at NERSC.  The colored curves with squares show the wall-time taken (in seconds) to compute the $a_{\ell m}$ up to various $\ell_{\rm max}$ values for $N_{\rm pnt}$ points on the sphere with the number of spline points fixed at $N_x=2048$ (see Appendix \ref{app:interpolation}).  The black dashed line shows the measured cost of determining what interval of the interpolation grid $\cos\theta$ falls into (i.e.\ `binning') and sorting the data in such a way that their corresponding grid values are arranged in ascending order.  The sorting scales as $\mathcal{O}(N\log N)$, as expected.  We see that this sorting, which only needs to happen once per dataset, dominates the cost for large $N_{\rm pnt}$.  It could be skipped if the data were already sorted, in which case the time would be dominated by binning (the black dotted line). }
\label{fig:scaling}
\end{figure}

Note that by following FKP we don't need to divide by the mean density, and thus the harmonic transforms can be done directly at the level of the data and random fields. If we are dealing with discrete objects, we don't really need to go through the pixelization step, and in fact there are some advantages to avoiding it.  For a set of point-like objects at $\{\theta_i,\phi_i\}$ with weights $\omega_i$, the spherical harmonic coefficients are
\begin{equation}
    a_{\ell m} = \sum_{i} \omega_i Y_{\ell m}^{\star}(\theta_i,\phi_i)\,,
\label{eqn:alm_def}
\end{equation}
where as usual
\begin{equation}
    Y_{\ell m}(\theta,\phi) = N_{\ell m} P_\ell^m(\cos\theta)e^{im\phi}
    \quad \text{and} \quad
    Y_{\ell,-m} = (-1)^m Y_{\ell m}^{\star}\,,
\end{equation}
with $P_{\ell}^{m}$ denoting the associated Legendre functions \cite{Lebedev72,Edmonds96} and with normalization
\begin{equation}
  N_{\ell m}= \sqrt{\frac{(2\ell+1)}{4\pi}\ \frac{(\ell-m)!}{(\ell+m)!}}
  \quad\mathrm{for}\ m\ge 0
\end{equation}
independent of position.  The computation of the exponential factor (sines and cosines) is straightforward, but calculating $P_\ell^m$ requires more attention.  This is handled in Appendix \ref{app:Ylm}, where we describe how to compute $Y_{\ell m}(\theta_k,0)$ and its first derivative for a grid of $\theta_k$, and in Appendix \ref{app:interpolation}, where we show how to efficiently interpolate from this grid using Hermite splines (which express the interpolated value as a linear combination of the function and its first derivative at the endpoints of the sub-interval). Storage and computation can be trivially distributed across devices, though this may not be necessary given how fast the algorithm is. With a CPU implementation in C it takes $\sim 30$ seconds to evaluate the SHT of $2^{24}\simeq 17\,$M points up to $\ell_{\rm max}\simeq 10^3$ on the Perlmutter computer at NERSC (using one AMD Milan CPU).  However the algorithm is well-suited for acceleration on GPUs, and using the JAX\footnote{https://github.com/google/jax} library on a Perlmutter node with four NVIDIA A100 GPU cores, we are able to obtain the SHT of $10^8$ points up to $\ell_{\rm max}\simeq 10^3$ in approximately the same amount of time ($\sim 30$ seconds; Fig.~\ref{fig:scaling}).  This establishes the direct SHT as feasible (and competitive with other approaches) and is likely sufficient for many purposes.  However, if high efficiency is needed, there are highly optimized libraries\footnote{As one example, the \href{https://gitlab.mpcdf.mpg.de/mtr/ducc/-/tree/ducc0}{ducc0 library} that is used in ref.~\cite{Reinecke23} has an ``adjoint\_synthesis\_general'' method that does the forward SHT.} that perform even better than this.

\subsection{Algorithm}
\label{sec:algorithm_summary}
Let us now summarize our proposed algorithm. Suppose we are given a set of data and randoms together with their corresponding weights.  Use Eq.~(\ref{eqn:alm_def}) on the data to compute $a_{\ell m}^{(d)}$ and on the randoms to produce $a_{\ell m}^{(r)}$, normalizing the randoms such that $a_{00}^{(r)}=a_{00}^{(d)}$.  Use Eq.~(\ref{eqn:wldef}) with $w_{\ell m}=a_{\ell m}^{(r)}$ to compute $W_{\ell}$ (subtracting shot noise) and Eq.~(\ref{eqn:mode_coupling}) to define $\widetilde{C}_b$, the binned, mode-decoupled, pseudo-$C_\ell$ estimate of the power spectrum (or, alternatively, the binned, normalized pseudo-$C_\ell$ estimate).

In the case that the mask or window is provided in pixelized form, the SHT should be done using whatever routines apply to that pixelization.  For example, for \textsc{HealPix} this can be done with the \texttt{anafast} routine of \texttt{healpy} \cite{healpy}.  Use Eq.~(\ref{eqn:alm_def}) on the data to compute $a_{\ell m}^{(d)}$.  Normalize the $a_{\ell m}^{(r)}$ such that $a_{00}^{(r)}=a_{00}^{(d)}$, i.e.
\begin{equation}
    \sum_{\rm pix} \Omega_{\rm pix} w_{\rm pix} = \sum_{i=1}^{N_d} \omega_i^{(d)}\,,
\end{equation}
where $\Omega_{\rm pix}$ is the area of the pixel, $w_{\rm pix}$ is the value of the mask in that pixel and $\omega_i^{(d)}$ are the weights for the data points.  If the mask is binary, i.e.\ $w_{\rm pix}=0$ or $1$, then the sum is $4\pi\,f_{\rm sky}$, where $f_{\rm sky}$ is the observed sky fraction.  Proceed as above to define $W_\ell$, $M$, $\mathcal{M}$ and $\widetilde{C}_b$.

\section{Examples}
\label{sec:examples}

\subsection{Golden spiral}

\begin{figure}
    \centering
    \resizebox{\columnwidth}{!}{\includegraphics{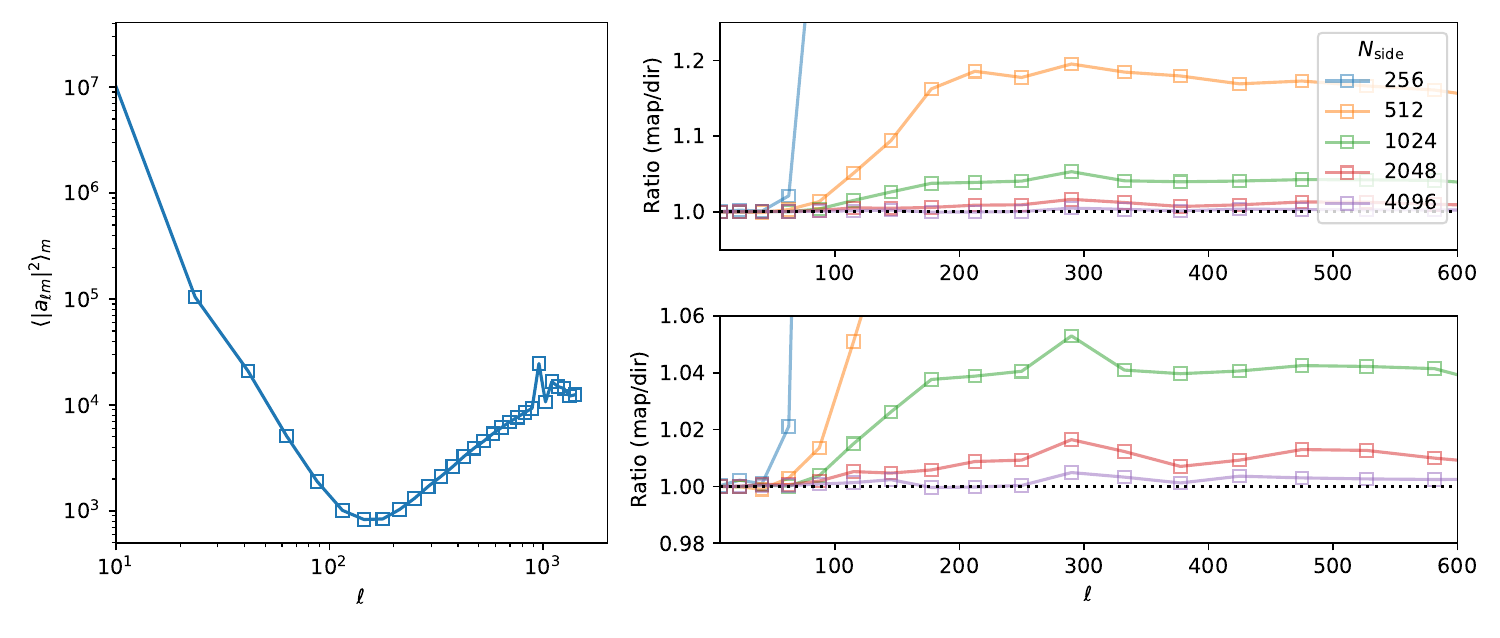}}
    \caption{Comparison of direct harmonic transform and transform via pixelization for the Golden Spiral example (see text).  Aliasing of small-scale power is particularly strong for this example. (Left) The average over $m$ of $|a_{\ell m}|^2$ computed by direct transform, in bins of $\Delta\ell\propto\sqrt{\ell}$.  (Top right) The ratio of the Healpix map-based power to the direct harmonic transform (i.e.~the left panel) on an expanded $y$-axis scale for $N_{\rm side}$ from 256 to 4096.  (Lower right) The same ratio on a smaller $y$-axis scale about unity. }
\label{fig:golden}
\end{figure}

Now we give two examples of the code and approach.  First we give an example with a lot of small-scale power to show how taming aliasing can require very fine pixelization.  For this example we consider a set of $N_{\rm pnt}=163\,840$ points laid out in a ``golden spiral'' (a.k.a.\ Fibonacci spiral) pattern between $\cos\theta=-1/2$ and $+1/2$:
\begin{equation}
    \cos\theta_k = \frac{1}{2}\left[ 1 - \frac{2k+1}{N_{\rm pnt}}\right]
    \quad , \quad
    \phi = \frac{2\pi k}{\varphi}
    \quad\mathrm{for}\quad
    0\le k\le N_{\rm pnt}
\end{equation}
where $\varphi=(1+\sqrt{5})/2$ is the golden ratio. This leads to an approximately uniform distribution of points in the covered region, and we perturb each point randomly by a very small amount so that the spacing isn't completely regular.  The left panel of Fig.~\ref{fig:golden} shows the $a_{\ell m}$ computed by direct harmonic transform, squared and averaged over $m$.  Note that this point set has significant `power' at high $\ell$ (analogous to a spike of power at the Nyquist frequency of a regular, Cartesian grid of points).  We then generate a \textsc{HealPix} map of these points with various $N_{\rm side}$ from 256 to 4096 and use the healpy \cite{healpy} ``\texttt{anafast}'' routine to compute the same average over $m$ of $|a_{\ell m}|^2$.  The right panels show the ratio of this estimate to the direct-sum approach (on a fine and expanded scale) up to $\ell=600$.  Note that the large amount of power at very high $\ell$ aliases into lower $\ell$ power unless $N_{\rm side}$ is quite large.  All cases shown have less than one point per pixel (in the `survey region') on average but still show significant excess power, even at $\ell$ of a few hundred.  If we do not perturb the points in the spiral from their initial positions the aliasing effect is even more extreme and none of the pixelized approaches converge for even moderate $\ell$.

\subsection{Lognormal mock catalog}

\begin{figure}
    \centering
    \resizebox{\columnwidth}{!}{\includegraphics{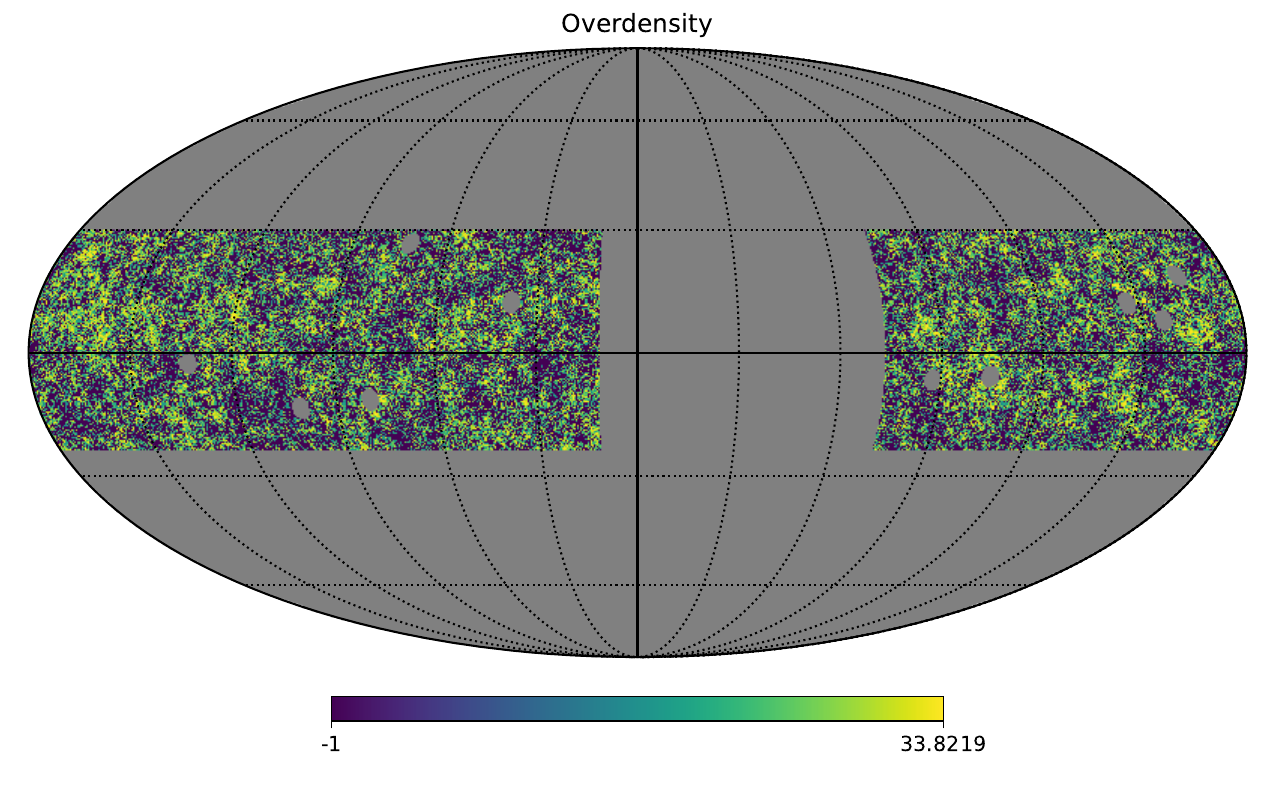}}
    \caption{Map of the overdensity field from our lognormal mock catalog, in Mollweide projection, covering approximately $34\%$ of the sky.  The catalog contains about $1$ million objects within the footprint shown, or $70\,\mathrm{deg}^{-2}$.  Given our fiducial clustering model the objects are signal dominated over most of the angular scales of interest to us, so the power spectrum is ``red''. }
\label{fig:map}
\end{figure}

The golden spiral example has a much bluer spectrum (i.e., with more small-scale structure) than we commonly consider in cosmology, and thus serves to illustrate the impact of aliasing.  Our second example is closer to the most common use-case for this sort of analysis in cosmology -- a set of points with an angular power spectrum that is approximately a (red) power-law, plus shot noise.  For this example we shall err in the other direction and enhance the clustering signal over the shot noise (e.g.\ as for galaxies in a narrow shell in redshift where the clustering is not washed out by projection) to produce a ``red'' spectrum with $C_\ell\approx\ell^{-3/2}$ that is sample variance limited to $\ell\approx 10^3$.  This should optimize the performance of the map-based approach.  To create such a set of points we first generate a Gaussian random field over the whole sphere at \texttt{Nside}=8192 using the \texttt{synfast} routine of \texttt{healpy} \cite{healpy}.  We exponentiate this field and then populate each pixel with probability proportional to the pixel value, perturbing the positions of points away from the pixel centers by a very small amount.  We then cut the set to $-0.4 < \cos\theta < 0.5$ and $0.2\le \phi < 5$, approximately one third of the sky, as an example of a ``survey mask'', and we place 10 `masked' regions of $2.5^\circ$ radius randomly within the footprint as a mockup of a (very) bright star mask. The resulting data, along with randoms thrown evenly within the same range, forms our mock dataset -- see Fig.~\ref{fig:map}.  We analyze this dataset using the direct approach and the more traditional approach of first forming the overdensity field on a \textsc{HealPix} grid.  In the absence of numerical effects both approaches should give the same pseudo-$C_\ell$, mode coupling matrix, etc.

\begin{figure}
    \centering
    \resizebox{\columnwidth}{!}{\includegraphics{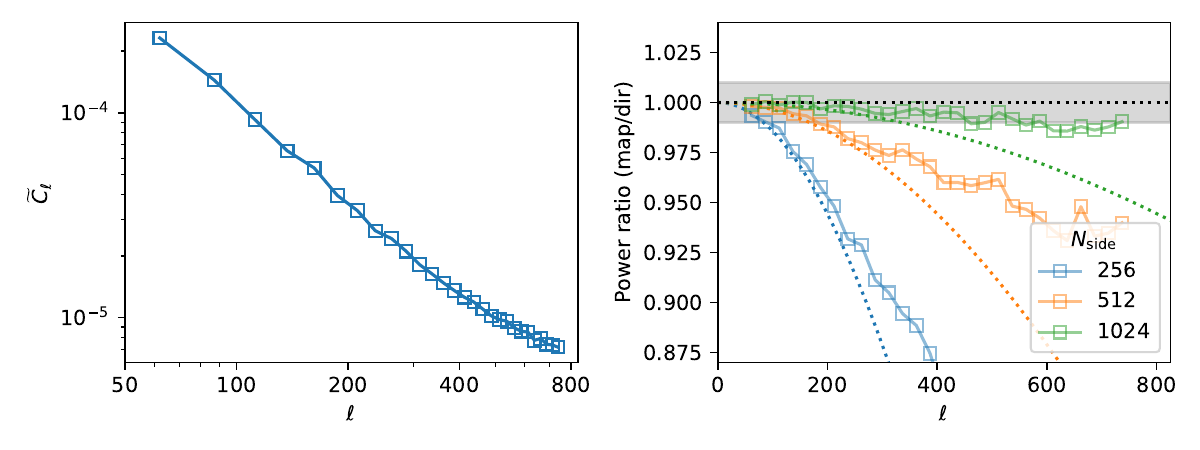}}
    \caption{A comparison of pseudo-spectra for the 1M mock galaxies shown in Fig.~\ref{fig:map}.  (Left) The binned and mode-decoupled angular power spectrum produced by direct spherical harmonic transform.  (Right) The ratio of the map-based estimates of the power spectrum to the direct (harmonic) spectrum for $N_{\rm side}=256$, 512 and 1024 ($N_{\rm pix}\approx 0.8$M, 3.1M, 12.5M).  The grey, horizontal band shows $\pm 1\%$ while the colored, dotted lines show the pixel window function (squared).  }
\label{fig:compare}
\end{figure}

We find that the `direct' approach is relatively robust to our analysis choices, and we get good results if the number of randoms exceeds the number of data by a factor of $20-50\times$.  The precise number of randoms needed will depend upon the degree of structure in the mask.  For the pixelized approach more care needs to be taken, especially when the overdensity field is generated by dividing a data map by a random map.  In this case we found it necessary to ensure tens of randoms per pixel on average to get robust answers (since ``division by $R$'' enhances the tails of the distribution).  However, for a very large number of random points and sufficiently high resolution, the agreement between the two methods was very good, as illustrated in Fig.~\ref{fig:compare}.  We note that if percent level convergence is desired for this sample one needs more stringent requirements than advocated in e.g.\  ref.~\cite{Leistedt13}.

\section{Conclusions}
\label{sec:conclusions}

Harmonic analysis of the clustering of point sources on the sphere can be computationally demanding with the traditional approach of generating pixelized maps that are then transformed.  We have presented an alternative, aimed specifically at analysis of cosmology surveys which cover significant fractions of the sky with the information mostly confined to large angular scales (due to observational and modeling limitations and the inherent complexities of galaxy formation that obscure cosmological signals). These include surveys such as DESI \cite{DESI}, Euclid \cite{laureijs2011euclid,EUCLID18}, SPHEREx \cite{Dore14}, LSST on the Vera Rubin Observatory \cite{LSST}, Roman \cite{Roman}, and their cross-correlations with the Simons Observatory \cite{SimonsObs}, the South Pole Observatory and CMB-S4 \cite{CMBS4}. The method directly transforms the point set as a sum of spherical harmonics and we show that this can be efficiently performed numerically and is well-suited to GPU acceleration.  In order to avoid having to divide by a mean density (e.g.\ estimated from a random catalog), we follow the approach more commonly adopted in 3D large-scale structure studies \cite{Feldman94} which involves forming $\bar{n}\,\delta$ by subtracting a scaled random catalog from the data: $n_g-\alpha n_r$.  If this field is transformed and squared one obtains the convolution of the angular power spectrum by a window function.  By applying a ``bandpower decoupling'' step, as usually done in CMB analyses \cite{Hivon02}, one obtains a properly normalized, binned estimate of the angular power spectrum.  This method is mathematically equivalent to the more common approach of forming an overdensity map that is then multiplied\footnote{If the mask is either 0 or 1 this multiplication step can be skipped, but in the more usual case of an apodized mask it needs to be included.} by a mask or window map and transformed, but more straightforward and robust.  We also discuss an alternative to ``bandpower decoupling'' which is again closer to the technique adopted in large-scale structure studies \cite{Feldman94}, where an overall normalization is applied without attempting to correct for the shape change induced by convolution with the mask.  We show that both approaches lead to very similar pseudo-spectra (Fig.~\ref{fig:conventions}).

Section \ref{sec:algorithm} introduces the algorithm and shows the connection to the ``FKP method'' \cite{Feldman94} usually employed in redshift surveys and the ``pseudo-$C_\ell$'' approach \cite{Hivon02} that is standard in CMB analyses.  We show how the measured pseudo-$C_\ell$ is related to the `true' spectrum on average and describe how our algorithm can be implemented (with some mathematical details relegated to appendices \ref{app:Ylm}, \ref{app:interpolation} and \ref{app:3j}).  Our GPU implementation is able to handle the SHT of $>10^8$ points in about 30 seconds, and can be trivially parallelized. 

Section \ref{sec:examples} presents some examples of the approach and a comparison to the more traditional route.  First we show how the aliasing that inevitably follows from pixelization can be challenging to handle if the spectrum under consideration has significant small-scale power.  Fig.~\ref{fig:golden} presents an extreme example in which the points being analyzed are laid out in a ``golden spiral'' pattern.  Fig.~\ref{fig:compare} shows a comparison of our method to the traditional approach for a spectrum closer to those considered in large-scale structure.  In this example we have generated a lognormal mock catalog (Fig.~\ref{fig:map}) with an enhanced clustering component in order to compare with a ``red'' spectrum which is most favorable to the pixel or map-based approach.  We see that for large enough $N_{\rm pix}$ the map-based approach converges to the direct SHT, though typically this requires $N_{\rm pix}>N_{\rm pnt}$ even for modest $\ell$.  In the oft-used approach where an overdensity field is generated by dividing the data map by the random map, the number of randoms should be $\gg 1$ per pixel so that the statistics of $1/R$ are not affected by the tails.  This, combined with the need for $N_{\rm pix}>N_{\rm pnt}$, typically requires $N_{\rm rand}\gg N_{\rm pnt}$.  Handling a large number of randoms and computing their pixel numbers can grow to be a non-trivial task, in addition to the $\mathcal{O}(N_{\rm pix}^{3/2})$ scaling of the basic algorithm.  Even when the use of randoms is avoided entirely, treating the `mask' or `weight map' linearly rather than dividing by it makes our algorithm more robust.

Finally we mention that giving the spherical harmonic transforms of the data ($a_{\ell m}$) and random ($w_{\ell m}$) fields provides a convenient complement to publishing a catalog of positions and weights.  The coefficients provide a compact represenation of the large-scale distribution, can be easily transformed between frames and allow simple computation of the 2-point and higher-point auto- and cross-spectra with other fields defined in harmonic space.  Handling systematics weights or mode deprojection is also straightforward (see e.g.\ the discussion in ref.~\cite{Alonso19}).  Harmonic-space apodization of the $a_{\ell m}$, e.g.\ as an alterative to higher-order charge assignment schemes, can also be applied to produce a band-limited function that can be analyzed in a map-based manner while leaving the large-scale properties of the signal unchanged (or changed in a known and controlled manner).

Our approach has several natural generalizations.  First, a `flat-sky' approximation to our method would simply entail replacing the $Y_{\ell m}(\theta, \phi)$ in equation~\eqref{eqn:alm_def} with $\exp[i \bm{\ell}\cdot\bm{x} ]$, where $\bm{x}$ is a vector living on the plane tangent to the sphere and $\bm{\ell}$ is the flat-sky analog of $(\ell,m)$. Since exponentials can be computed very efficiently in terms of sines and cosines, it may be possible to obtain the Fourier coefficients exactly, even for a large number of points, by evaluating the exponential at all the points rather than interpolating. However, given the accuracy and speed of our full-sky implementation, this approximation is likely unnecessary. In fact, our method provides a straightforward and efficient way to upgrade analyses to the full-sky formalism.

Second, though we have presented the algorithm for describing the (scalar) density field, the generalization to fields with non-zero spin (such as cosmic shear) is relatively straightforward: rather than use $Y_{\ell m}$, one uses spin-spherical harmonics (${}_{\pm 2}Y_{\ell m})$, which can also be efficiently computed using recurrence relations.  One can also use this approach to implement a ``radial-angular'' (e.g.\ spherical Fourier-Bessel) analysis of 3D fields \cite{Lahav94,Fisher94,Heavens95,Castorina18,Grasshorn21,Gao23}, such as galaxy redshift surveys.  Such an approach would also be particularly appropriate for harmonic analyses of the Ly$\alpha$ forest, where sightlines are sparse but each sightline is well sampled in the radial direction, as an alternative to the 3D flux power spectrum \cite{Font18,Karim23}.  We choose not to pursue these directions here.

We make our \texttt{DirectSHT} code publicly available\footnote{\url{https://github.com/martinjameswhite/directsht/tree/main}} and provide several tutorial notebooks.

\acknowledgments
We are grateful to Noah Sailer, Julien Carron, Zvonimir Vlah and David Alonso for useful comments on our manuscript. A.B.L. would also like to thank Marcelo Alvarez, Shamik Ghosh, Minas Karamanis, David Nabergoj and Stephen Bailey for discussions regarding the JAX library.
M.W.~is supported by the DOE.
This research has made use of NASA's Astrophysics Data System, the arXiv preprint server, the Python programming language and packages \textsc{NumPy, Matplotlib, SciPy, AstroPy, JAX}, \textsc{HealPy}~\cite{healpy}.
This research is supported by the Director, Office of Science, Office of High Energy Physics of the U.S. Department of Energy under Contract No. DE-AC02-05CH11231, and by the National Energy Research Scientific Computing Center, a DOE Office of Science User Facility under the same contract.

\appendix

\section{Numerical and implementation details}

\subsection{Computing the spherical harmonics}
\label{app:Ylm}

In order to compute our harmonic coefficients we need to be able to (efficiently) compute $Y_{\ell m}(x=\cos\theta,\phi)$.  It suffices to evaluate these for $\ell\ge 0$, $0\le m\le\ell$ and $x\in [0,1)$.  The values for $m<0$ and $x<0$ can be computed by symmetry since $Y_{\ell,-m}=(-1)^m Y_{\ell m}^\star$ while reflection in $x$ incurs a phase $(-1)^{\ell -m}$.  Let us decompose $Y_{\ell m}$ as
\begin{equation}
    Y_{\ell m}(\theta,\phi) = \sqrt{\frac{(2\ell+1)}{4\pi}}\ \bar{P}_\ell^m(x=\cos\theta)\, e^{im\phi}
    \quad , \quad m\ge 0
\end{equation}
where
\begin{equation}
    \bar{P}_\ell^m(x) \equiv \sqrt{\frac{(\ell-m)!}{(\ell+m)!}}\ P_\ell^m(x)
    \quad , \quad m\ge 0
\end{equation}
is a normalized version of the associated Legendre functions, with the normalization included for numerical convenience (since $P_\ell^m$ can become very large for $m\approx\ell$).  The only computationally challenging piece is $\bar{P}_\ell^m$.  However, these functions are smooth so their values and first derivatives can be computed on a grid of $x=\cos\theta$ and then interpolated using a Hermite spline.  Both $\bar{P}$ and its derivative can be computed very efficiently using recurrence relations\footnote{See e.g.\ https://dlmf.nist.gov/14.10} that can be derived from their generating function \cite{Lebedev72,Edmonds96}, so the resulting algorithm is fast and highly vectorizable.

\begin{figure}
    \centering
    \resizebox{\columnwidth}{!}{\includegraphics{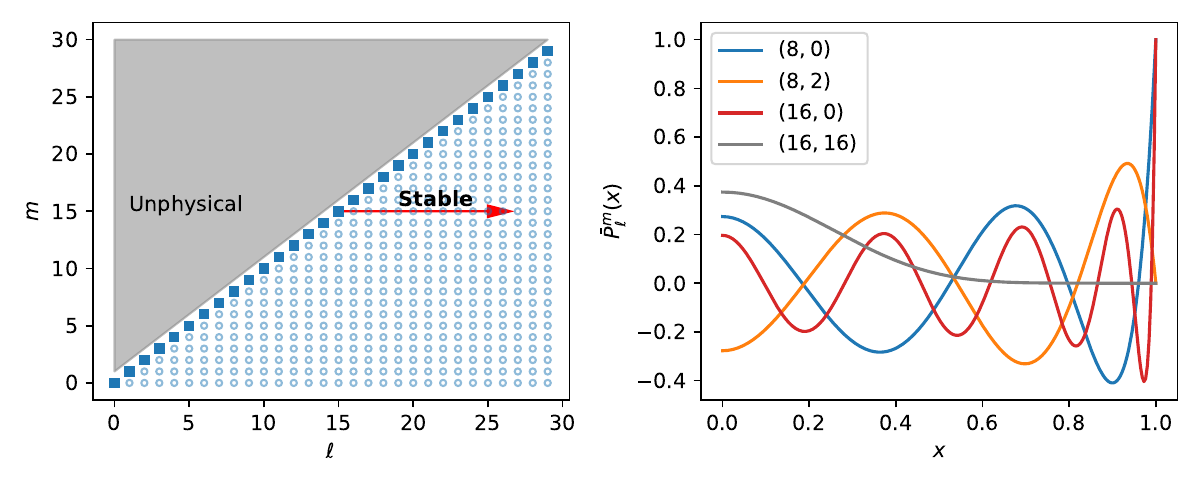}}
    \caption{(Left) An illustration of the method for computing the (scaled) Legendre functions, $\bar{P}_\ell^m(x)$ via recurrence.  $\bar{P}$ is dominant on the degree ($\ell$) so the recurrence relations are stable with increasing $\ell$ at fixed $m$. (Right) Some examples of $\bar{P}_\ell^m(x)$ for several $(\ell,m)$ pairs. }
\label{fig:legendre_method}
\end{figure}

The $\bar{P}$ can be computed by recurrence starting from
\begin{equation}
    \bar{P}_m^m = -\sqrt{1-x^2} \sqrt{ 1-(2m)^{-1} } \ \bar{P}_{m-1}^{m-1}
    \quad \text{and} \quad
    \bar{P}_{m+1}^m = \sqrt{2m+1}\, x\ \bar{P}_m^m\,,
\end{equation}
with the special cases $\bar{P}_0^0(x)=P_0(x)=1$ and $\bar{P}_1^0(x)=P_1(x)=x$.  Now, $\bar{P}$ is dominant on the degree ($\ell$) and minimal on the order ($m$), so for all other $\ell$ and $m$
\begin{equation}
    \bar{P}_\ell^m
    = \sqrt{\frac{(\ell-m)}{(\ell+m)}}\, \left[ \frac{2\ell-1}{\ell-m}\ x\,  \bar{P}_{\ell-1}^m(x) -  \sqrt{\frac{(\ell-m-1)}{(\ell+m-1)}} \, \frac{\ell+m-1}{\ell-m}\  \bar{P}_{\ell-2}^m(x) \right]
\end{equation}
is stable with $m$ increasing from $0$ and $\ell$ increasing from $m+2$ for each $m$ (see Fig.~\ref{fig:legendre_method}).  If high numerical efficiency is desired, there are reformulations of this recurrence that require fewer floating point operations \cite{Ishioka18}.

Once the $\bar{P}_\ell^m$ are known, the derivatives can be computed from
\begin{equation}
    \frac{d}{dx} \bar{P}_\ell^m(x) = -\frac{\ell\,x}{1-x^2} \bar{P}_\ell^m(x)
    + \frac{\ell+m}{1-x^2} \sqrt{\frac{(\ell-m)}{(\ell+m)}} \bar{P}_{\ell-1}^{m}(x)
\end{equation}
for $\ell>0$, with $P_0^\prime(x)=0$ and $P_1^\prime(x)=1$.

\subsection{Interpolation}
\label{app:interpolation} 

The sum in Eq.~(\ref{eqn:alm_def}) must be performed for both the data and random samples.  It is not uncommon for this to involve $10^7$ or $10^8$ points, and while the problem is ``embarrassingly parallel'', some care is still required in this evaluation.  Since $Y_{\ell m}(\theta,\phi)$ is smooth, it is computationally advantageous to compute $Y_{\ell m}$ only at certain sample points and then interpolate between them. On the other hand, the $\phi$-dependence can be included exactly at every $\phi_r$ position, as it is cheap to obtain. For notational convenience, let us define $y_{\ell m}(x)\equiv Y^{\star}_{\ell m}(\theta,0) = Y_{\ell m}(\theta,0)$ with $x=\cos\theta$.

We found that when $N_{\rm pnt}$ is large the interpolation step can become a bottleneck for the calculation unless some care is taken. Thankfully, our analytic scheme to calculate $\bar{P}_{\ell m}(x)$ and its derivatives (appendix~\ref{app:Ylm}) allows for fast and accurate interpolation using cubic Hermite splines, which are the unique piecewise-cubic polynomial with the correct function value and first derivative at the sample points. For a set of $N_x$ points labeled $x_j$ where our function is sampled, and defining $t_j \equiv (x - x_j)/(x_{j+1}-x_{j})$, these splines are
\begin{align}
    S^{(\ell,m)}(x) = \sum_{j=0}^{N_x-1} \big[
    & c_{j 0}(x) y_{\ell m}(x_j) + c_{j 1}(x) (x_{j+1}-x_{j}) y'_{\ell m}(x_j) \nonumber \\
    & + c_{j 2}(x) y_{\ell m}(x_{j+1}) + c_{j 3}(x) (x_{j+1}-x_{j}) y'_{\ell m}(x_{j+1})\big]\,,
\end{align}
with
\begin{equation}
      c_{j 0}(x) \equiv (2t_j+1)(1-t_j)^2 \ , 
    \ c_{j 1}(x) \equiv t_j (1-t_j)^2 \ , 
    \ c_{j 2}(x) \equiv t_j^2 (3-2t_j) \ , 
    \ c_{j 3}(x) \equiv t_j^2 (t_j-1) 
\end{equation}
if $x_j \leq x < x_{j+1}$ and are zero otherwise. Though the splines $S^{(\ell,m)}(x)$ are unique to each $(\ell,m)$ pair, the dependence enters only through the values of $y_{\ell m}$ and its derivative at the sample points -- in which the spline is linear. Both of these can be tabulated in advance following appendix~\ref{app:Ylm}, thus eliminating the need to re-compute the interpolation weights at each $(\ell,m)$, which is an expensive requirement of some other schemes. In addition, this being a `local' interpolation method linear in $y$ and $y^\prime$, we are free to choose our sample points without regard for the ringing artefacts that plague `global' methods or that require careful sampling (e.g.\ using Chebyshev nodes) to ensure a linear system can be stably inverted.  In fact there is no linear system to solve for when computing the Hermite spline.

While the dependence on $(\ell,m)$ enters the splines only through the values of $y_{\ell m}$ and its derivative at the sample points, the $c_{j \alpha}$ depend upon $x$. This dichotomy is key to an efficient implementation. Interpolating with our cubic Hermite spline, we can write
\begin{align}
    a_{\ell m} & = \sum_{r=0}^{N_{\rm pnt}} w_r e^{-i m \phi_r} y_{\ell m}(x_r) \nonumber \\
    & = \sum_{r=0}^{N_{\rm pnt}} w_r e^{-i m \phi_r} \sum_{j=0}^{N_x} \big[c_{j 0}(x_r) y_{\ell m}(x_j) + c_{j 1}(x_r) y'_{\ell m}(x_j) \nonumber \\
    & \hphantom{=  \sum_{r=0}^{N_{\rm pnt}} w_r e^{-i m \phi_r} \sum_{j=0}^{N_x} \big[ }+ c_{j 2}(x_r) y_{\ell m}(x_{j+1}) + c_{j 3}(x_r) y'_{\ell m}(x_{j+1})  \big] \nonumber \\
    & =  \sum_{j=0}^{N_x} \big[  v_{j 0}^{(m)} y_{\ell m}(x_{j})  + v_{j 1}^{(m)} y'_{\ell m}(x_j)   + v_{j 2}^{(m)} y_{\ell m}(x_{j+1}) + v_{j 3}^{(m)} y'_{\ell m}(x_{j+1})   \big]
\end{align}
where we have defined
\begin{equation}
    v_{j \alpha}^{(m)} \equiv \sum_{r=0}^{N_{\rm pnt}} w_r e^{-i m \phi_r} c_{j \alpha}(x_r)  \,.
\end{equation}
These objects can be tabulated efficiently harnessing the fact that the $c_{j \alpha}(x_r)$ are zero outside of the spline region that $x_r$ falls into: we therefore just need to accumulate contributions within bins bounded by the $x_i$ sample points, rather than actually summing over all $N_{\rm pnt}\sim 10^6-10^8$ elements at each $j$. Since it needs to happen at every $m$, this binning is one of the most computationally demanding steps of our pipeline. It is ideally carried on a GPU, the high throughput of which allows for efficient parallelization of the accumulation step.

\begin{figure}
    \centering
    \resizebox{\columnwidth}{!}{\includegraphics{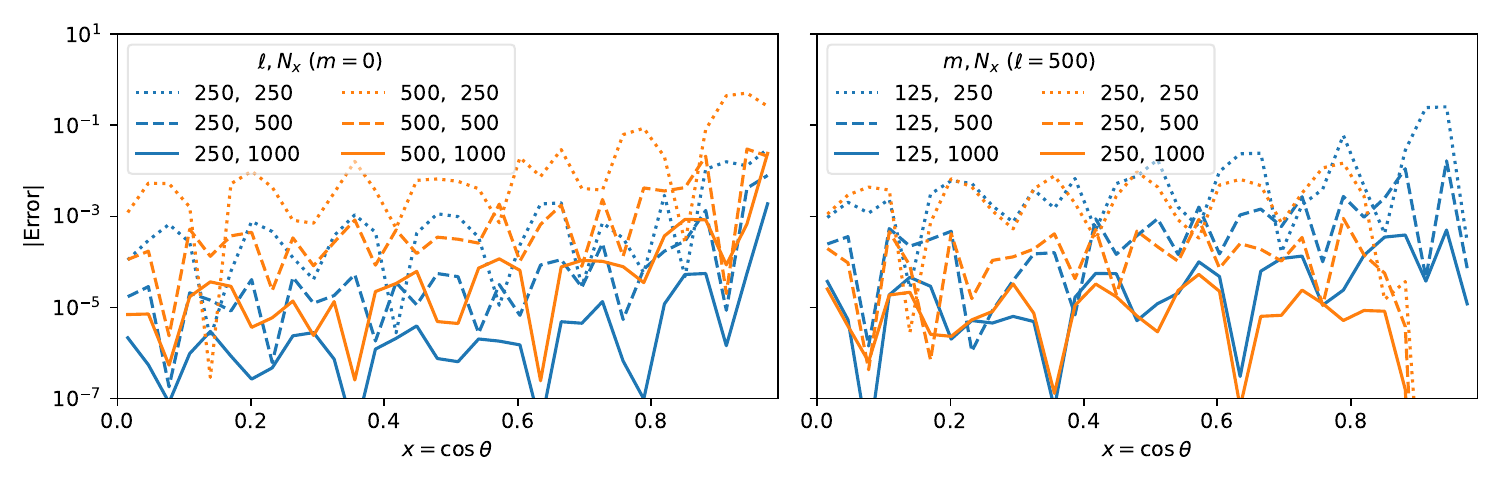}}
    \caption{The absolute error induced in our interpolation of $y_{\ell m}(x)$ as a function of $x=\cos\theta$.  (Left) The absolute error for $\ell=250$ and 500 with $m=0$, for different numbers of spline points.  (Right) The absolute error for $\ell=500$ and $m=125$ and 250 for different numbers of spline points (the error for $m=500$ is much smaller).  The characteristic size of $y_{\ell m}(x)$ is $\mathcal{O}(1)$ while for $N_x=2\ell_{\rm max}$ (solid lines) the error is $\mathcal{O}(10^{-4})$ except near $x\approx 1$ where for $m=0$ it grows rapidly (see text).}
\label{fig:interp_error}
\end{figure}

Fig.~\ref{fig:interp_error} shows the interpolation error as a function of $x=\cos\theta$ for several values of $(\ell,m)$ and numbers of spline points, $N_x$.  It appears that $N_x=N_\ell$ returns fractional errors of approximately $10^{-3}$ for $\ell=500$ and $m=0$ while $N_x=2N_\ell$ does about an order of magnitude better.  However, in practical situations involving many points, the implied error is actually less.  For example, we find that the fractional error on $W_\ell$ computed from the random catalogs associated with the mock galaxy footprint described in \S\ref{sec:examples} and shown in Fig.~\ref{fig:map} is less than one percent for $N_x=N_\ell/2$, and almost undetectable for $N_x=N_\ell$.  Furthermore, this fractional error occurs at $\ell\approx\ell_{\rm max}$, where $W_\ell$ is very small anyway.  Thus, choosing $N_x\ge N_\ell$ should lead to very small errors in practice.  Fortunately, the code is sufficiently fast (Fig.~\ref{fig:scaling}) that the choice of $N_x$ is not particularly critical.

Our interpolation scheme does have some drawbacks.  Near $x\approx 1$, the largest of the $y_{\ell m}$ are those with $m=0$. These become very steep as $x\to 1^{-}$ (Fig.~\ref{fig:legendre_method}).  In fact $P_\ell^\prime(1)=\ell(\ell+1)/2$, so the slope grows quadratically with the degree.  This renders any interpolation scheme ineffective for large enough $\ell$ (see e.g.\ Fig.~\ref{fig:interp_error}).  For our situation we are able to simply limit $|x|$ such that we avoid the poles, and the rest of the function is very well behaved. For a fully general scheme, one approach would be to split off the points where $|x|$ exceeds some tolerance and then explicitly compute $y_{\ell m}(x)$ for those points either by recurrence as in appendix~\ref{app:Ylm} or via one of the many small-angle approximations, e.g.\ $P_\ell^m(x)\approx J_m(\ell\theta)$, coupled with the ability to rapidly compute special functions \cite{Lebedev72}.  An alternative is to split the sample into regions near the equator and regions near both poles.  The equatorial points are treated as above.  The polar regions can be rotated to the equator, the SHT applied, and then the $a_{\ell m}$ rotated back to the original frame using Wigner matrices (which are block-diagonal in $\ell$).  We utilize the latter approach in our code for any points with $|x|>3/4$.  The Wigner matrices can also be computed from recurrence relations, see e.g.\ \S 3.3 of ref.~\cite{Price23}.

\section{Wigner 3j symbols}
\label{app:3j}

In computing the mode-coupling matrix (Eq.~\ref{eqn:chat}) it is necessary to have an efficient means for evaluating the Wigner $3j$ symbols.  This can also been done via recurrence \cite{Edmonds96} and we include the steps here for completeness.  Let the three multipoles be $\ell_1$, $\ell_2$ and $\ell_3$.  For $m_1=m_2=m_3=0$, the $3j$ symbol is invariant under any permutation of the $\ell_i$ and vanishes if $L=\ell_1+\ell_2+\ell_3$ is odd.  Non-vanishing values can be obtained from
\begin{equation}
    \begin{pmatrix} \ell_1 & \ell_2 & \ell_3 \\ 0&0&0 \end{pmatrix}
    = \sqrt{\frac{(L-2\ell_2-1)(L-2\ell_3+2)}{(L-2\ell_2)(L-2\ell_3+1)}}
    \ \begin{pmatrix} \ell_1 & \ell_2+1 & \ell_3-1 \\ 0&0&0 \end{pmatrix}\,,
\end{equation}
with the terminal case
\begin{equation}
    \begin{pmatrix} \ell & \ell & 0 \\ 0&0&0 \end{pmatrix}
    = \frac{(-1)^\ell}{\sqrt{2\ell+1}}
    \quad .
\end{equation}
This allows rapid evaluation of the required $3j$ symbols, and the calculation can be further optimized if previously computed values are cached.

\bibliographystyle{JHEP}
\bibliography{main} 
\end{document}